\documentstyle[12pt,blois,psfig]{article}

\newcommand{\subsun}{\mbox{$_{\odot}$}}
\newcommand{\sun}{\mbox{$\odot$}}

\def\ltsima{$\; \buildrel < \over \sim \;$}
\def\gtsima{$\; \buildrel > \over \sim \;$}
\def\simlt{\lower.5ex\hbox{\ltsima}}
\def\simgt{\lower.5ex\hbox{\gtsima}}

\def\arcm{$'~$}
\begin{document}
\heading{LYMAN BREAK GALAXIES AT $z \sim 3$ AND BEYOND} 

\author{C.C. Steidel$^{1}$, K.L. Adelberger$^{1}$, M. Dickinson$^{2}$,
M. Giavalisco$^{2}$, and M. Pettini$^{3}$}
{$^{1}$ Palomar Observatory, Caltech,
Pasadena, CA USA}{$^{2}$ Space Telescope Science Institute, Baltimore, MD, USA}{$^3$ Institute of Astronomy, Cambridge, UK} 

\begin{bloisabstract}
We report on the status of large surveys of photometrically selected 
star forming galaxies at $z \sim 3$ and $z \sim 4$, with particular
emphasis on both the advantages and the limitations of selecting objects using the
``Lyman break'' technique. Current results on the luminosity functions,
luminosity densities, color distribution, star formation rates, clustering properties, and
the differential evolution of the population as a function of redshift
are summarized. 
\end{bloisabstract}

\section{Introduction}

We have heard a great deal at this meeting about the exciting results on
high redshift galaxies that are beginning to be obtained at IR, sub-millimeter,
and radio wavelengths. These are clearly very special times for studies of
the distant universe, which are destined to unfold rapidly with the
coming of ubiquitous 8m-class telescopes, new observing techniques, capabilities
at previously unexplored wavelengths, 
and, not least, by increased awareness of what one should
be looking for to successfully find galaxies at high redshift. If any message
emerges when one looks at this meeting as a whole, it is that there are many complementary 
ways to acquire information on the distant universe, and one probably needs
some synthesis of {\it all} of the observations, at all wavelengths, to
appreciate the full picture of what is going on. 
It is important to be cognizant of 
the advantages and limitations of each technique, and to pay attention to what
is happening throughout the field and not fall victim to ``wavelength myopia'',
wherein one promotes observations at some particular wavelength 
without accounting for what one has learned, or can learn, from other methods. 
In the interest of
trying our best to acquire this kind of global view of the current
situation, in this short summary we make an attempt to give an honest assessment
of the virtues and limitations of studying high redshift galaxies by selecting
them using the ``Lyman break'' technique. 

Any high redshift object that produces UV continuum photons (whether it be
produced by star formation or by AGN activity) that manage to escape
the parent object without being absorbed and re--radiated in the far-IR
is capable of being detected using the Lyman break technique; 
the guaranteed spectral feature at the ionization edge of neutral hydrogen
at 912 \AA\ in the rest frame can be discerned easily with broad--band
photometry. While the idea of using this feature to find distant galaxies
is not new (see \cite{Meier76} for an amazingly prescient discussion of
the predicted appearance and possible search methods for ``primeval--galaxies''), 
it is only recently
that it has been successfully implemented \cite{SPH,S96,Madau96,Lowenthal97}.
In only a few years,  the high efficiency of the technique
has resulted in very rapid progress in accumulating large samples
of high redshift galaxies. Large samples allow statistical insight into the
nature of a particular galaxy population, and enable the same kind of detailed
analyses that have recently been completed at $z \sim 1$ (e.g. \cite{Lilly96,Ellis96,
Cowie96,Cohen96}.

Of course, Lyman break galaxies (LBGs) are certainly not the whole story, and
many of the talks at this meeting have emphasized how incomplete any assessment of
the progress of galaxy formation at high redshift can be if based entirely on
objects selected in the far--UV. The far--UV is easily subject to extinction 
by dust, and is sensitive essentially {\it only} to current (unobscured) massive star
formation, and not at all to accumulated stellar populations of a galaxy (as would
arguably be the case for near--IR selected samples at moderate redshift-- see, e.g., 
Peter Einsenhardt's contribution from this meeting.) Thus, one learns nothing
about the star formation history of individual objects by studying Lyman break galaxies (although one
is not obviously biased {\it against} objects that have substantial stellar mass, so
long as they also have current star formation, and it is possible to make measurements
sensitive to older stars once the objects are found) and it is not clear {\it a priori} to what
level dust has affected what appears or does not appear in the sample.

Having stated these {\it caveats} up front, we move to the advantages of LBG selection.
An obvious one is that the surface density of these high redshift galaxies is
high enough that large samples can be acquired using existing CCD imagers and
imaging spectrographs on 4m-10m telescopes. This has allowed the first
detailed analyses of the clustering properties of galaxies at $z>>1$ \cite{S98a,S98b,G98,A98}.
The clustering properties are a real acid test
of whether the basic ideas we have about how galaxies form are anywhere
close to correct. Even a field as small as the Hubble
Deep Field provides large enough samples of LBGs that many have used it
to infer the entire history of galaxy and star formation (e.g., \cite{Madau96,Sawicki97},)
although we will argue below that this application is 
premature.

\begin{figure}
\centering\mbox{\psfig{file=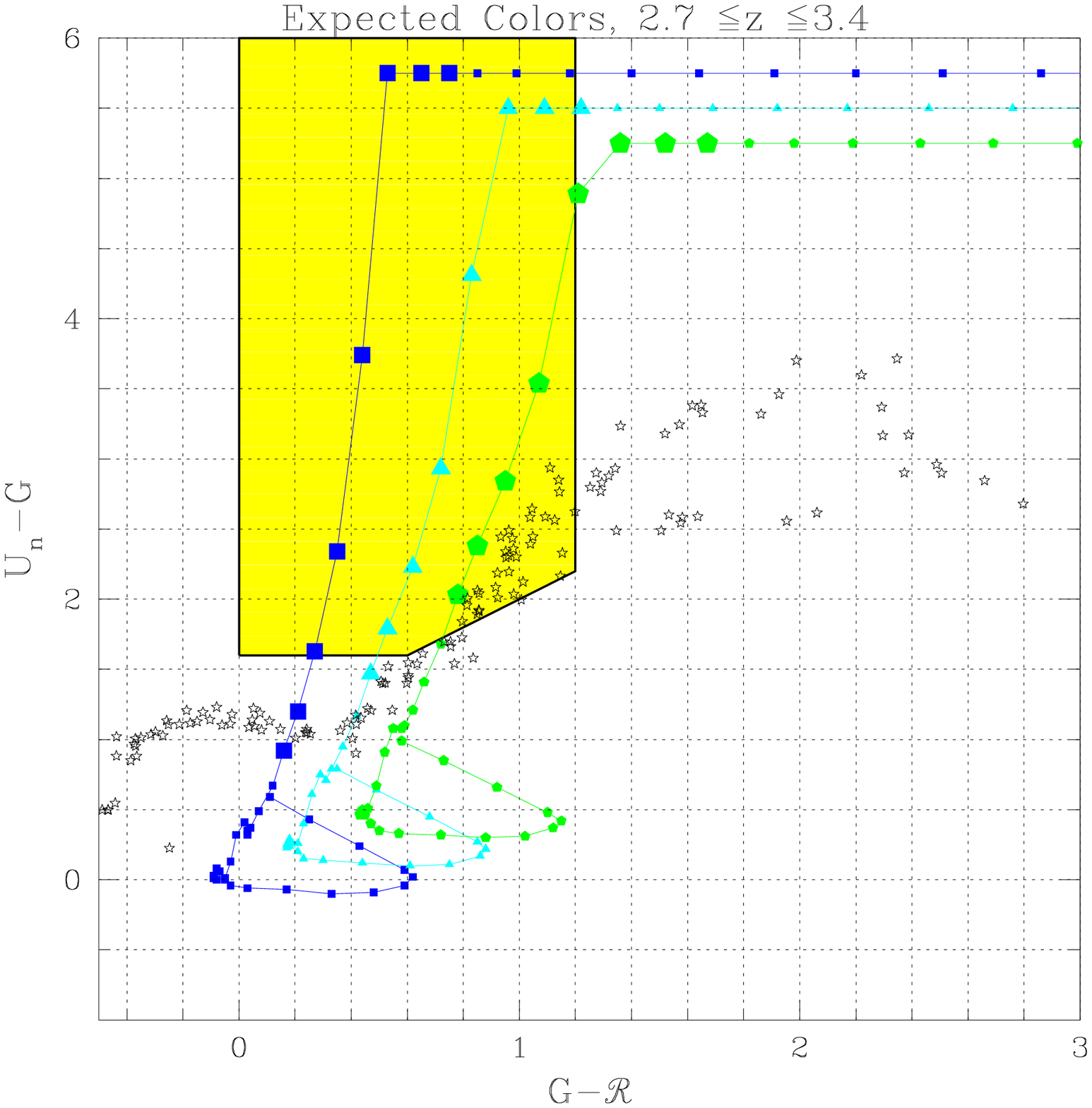,height=7cm}\psfig{file=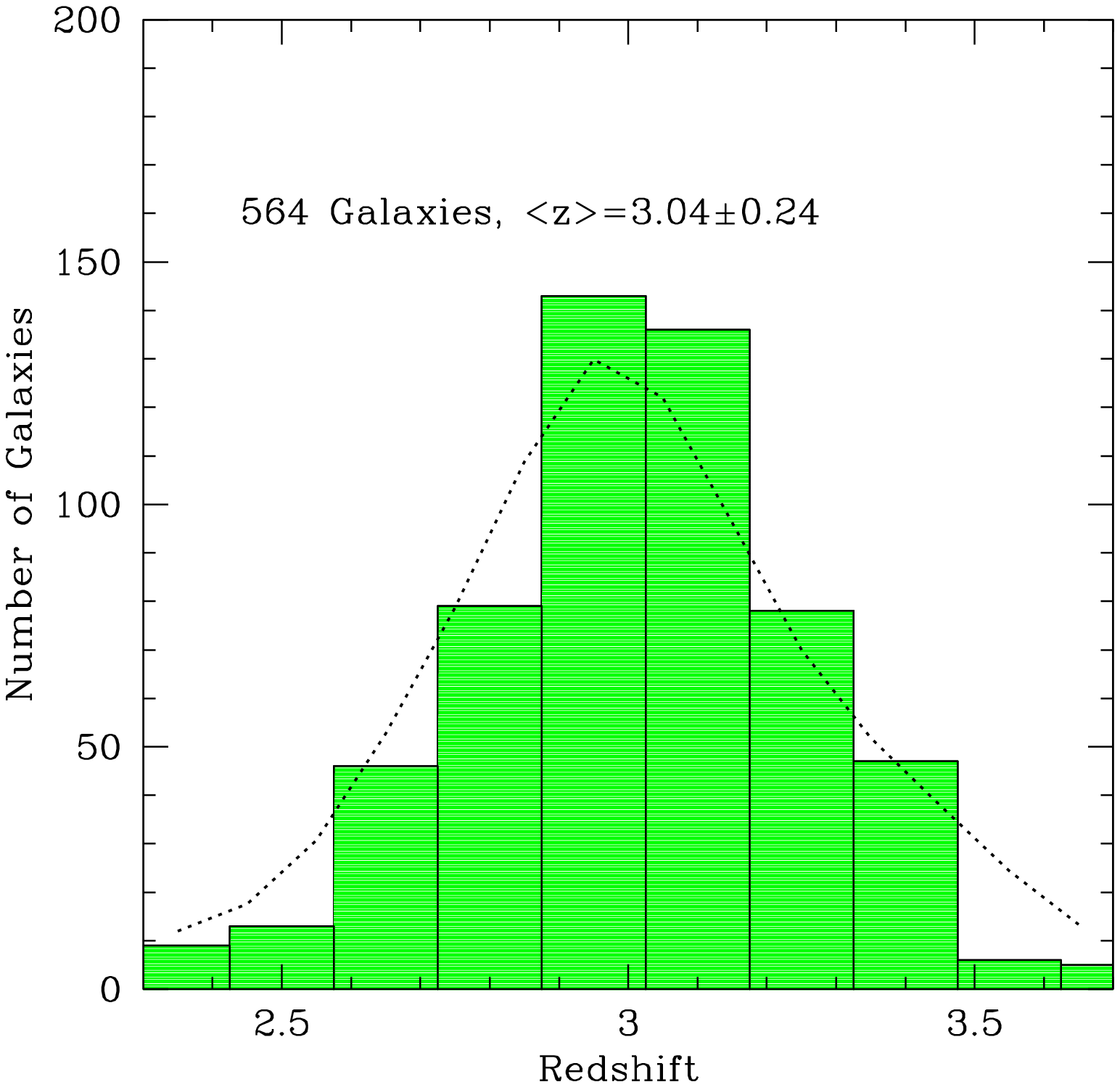,height=7cm}}
\caption[]{(Left) Theoretical 2--color diagram illustrating the region from
which $z \sim 3$ LBGs are selected. The 3 color tracks correspond to a model spectrum
subjected to different amounts of reddening, spanning the range observed for
most of the real galaxies: E(B-V)=0 (squares), E(B-V)=0.15 (triangles),
and E(B-V)=0.30 (pentagons). The points are at intervals of $\Delta z=0.1$, with
redshifts between 2.7 and 3.4 enlarged. 
(Right) The current redshift
histogram for objects selected from the indicated region of the two color diagram. The
dotted curve represent the predictions based on the models shown in the left panel, accounting
for known sources of incompleteness.  
 }
\end{figure}

One of the reasons for the success of these early LBG studies is
that the observed--frame optical (and therefore rest--frame UV  for the
high redshift objects) is still the spectral region in which the
greatest sensitivity is achievable from the ground. For example, the spectroscopic
surveys for LBGs that we will describe below achieve limiting sensitivities of
about 200 {\it nano}-Jy at 0.7 $\mu$m. With purely photometric techniques, particularly using
deep HST images, these numbers can be improved upon significantly. This large
dynamic range allows the detection (in principle) of heavily obscured high
redshift galaxies so long as there is a small leakage of UV photons. Put another way,
the ground--based LBG samples can detect the equivalent of $\sim$ several
$\times 10^{9}$ L\subsun\ at $z\sim 3$, so that an object like Arp 220 could
be detected even if only $\sim 1\%$ of the energy emerged in the far--UV.
Now, it might not be recognized as an ultra-luminous galaxy on the basis
of the far--UV observations, but there is some hope that 
one can disentangle the effects of extinction and
reddening for the LBGs, as discussed below.

\section{Lyman Break Galaxies at $z\sim 3$}

Because LBGs have been reviewed many times recently (e.g., \cite{D98,P97,S98b},
in this
short summary we concentrate on new results and on general inferences, rather
than on details.   

At the time of this writing, we have obtained spectroscopic redshifts for
more than 750 galaxies in the redshift range $2 \le z \le 5$, using various
color--selection criteria to choose targets for spectroscopy. The most
intensive work to date has been on a sample at $z \sim 3$, for which
about 0.25 square degrees (in 6 high latitude fields, mostly of
angular extent 9\arcm\ by 18\arcm) has been imaged (using the facilities of
many 4m--class telescopes, including Palomar, CTIO, KPNO, and WHT) deeply
enough to reliably select candidates to ${\cal R}_{AB}=25.5$. Our spectroscopic
sample has been obtained exclusively using the Keck telescopes and the
Low Resolution Imaging Spectrograph \cite{Oke95}. The selection criteria and
the current redshift histogram are shown in Figure 1.

Unlike many spectroscopic
surveys, the imaging portion for target selection is absolutely crucial and
is also very time--consuming: a typical field, which is imaged in
the $U_nG{\cal R}I$ filters to a uniform depth of $\sim 29-29.5$ 
mag per square arc second surface brightness limits, requires more than
2 full nights of good weather and good seeing on a 4m--class telescope.
Huge gains in this aspect of the project could be realized with the new
generation of wide--field mosaic imagers, tiled with CCDs having good UV/blue response.
As we will discuss below (\S 3), we have just completed a similarly large imaging
survey (but smaller spectroscopic sub--sample) of galaxies at $z \sim 4$, using
analogous selection criteria but different filter combinations for candidate
selection. 

\subsection{Far--UV Luminosity Function}

Recently we have undertaken a major effort to quantify the details of our
selection criteria, and to understand the various sources of incompleteness
in the ground--based color--selected samples [\cite{A99,S99}]. Extensive
Monte Carlo simulations have been used to determine the effects on our samples of 
photometric errors, blending with foreground objects, and ``template incompleteness''
(i.e., what fraction of all galaxies at $z \sim 3$ would be detected given
our color selection criteria). The results of these simulations provide an
estimate of the effective volume surveyed as a function of apparent magnitude
(essential for calculating a luminosity function) given the incompletenesses and an estimate
of the true distribution of galaxy colors at a given redshift (see \S 2.2). The latter
can then be used in conjunction with models to obtain a better estimate of the effective volume
surveyed for any set of LBG selection criteria, including those used in the HDF 
\cite{Madau96,MPD98,D98}. The resulting composite far--UV luminosity function is shown
in Figure 2a. 

Note that a Schechter \cite{Schech76} function is a very good fit to the distribution, and
that the quite--steep faint end slope is obtained either by using only
the ground--based data to ${\cal R}=25.5$, or by including the 
HDF data which reach to considerably fainter magnitudes. Most of the
difference between this LF and an earlier version presented in \cite{D98}, which
had a much flatter faint end slope, is attributable to the greatly--improved 
incompleteness corrections for the ground--based samples.  

\begin{figure}
\centering\mbox{\psfig{file=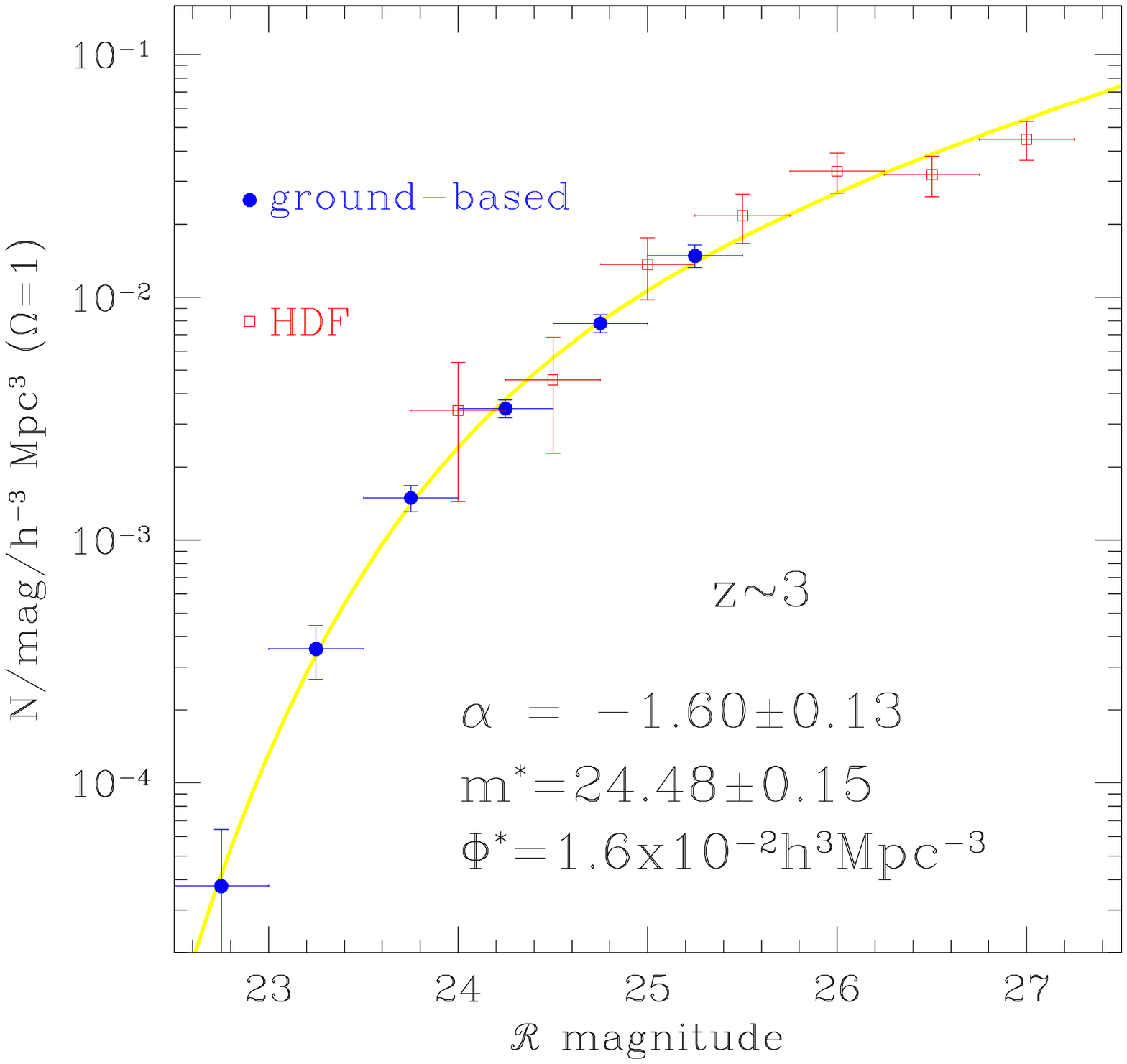,height=7cm}\psfig{file=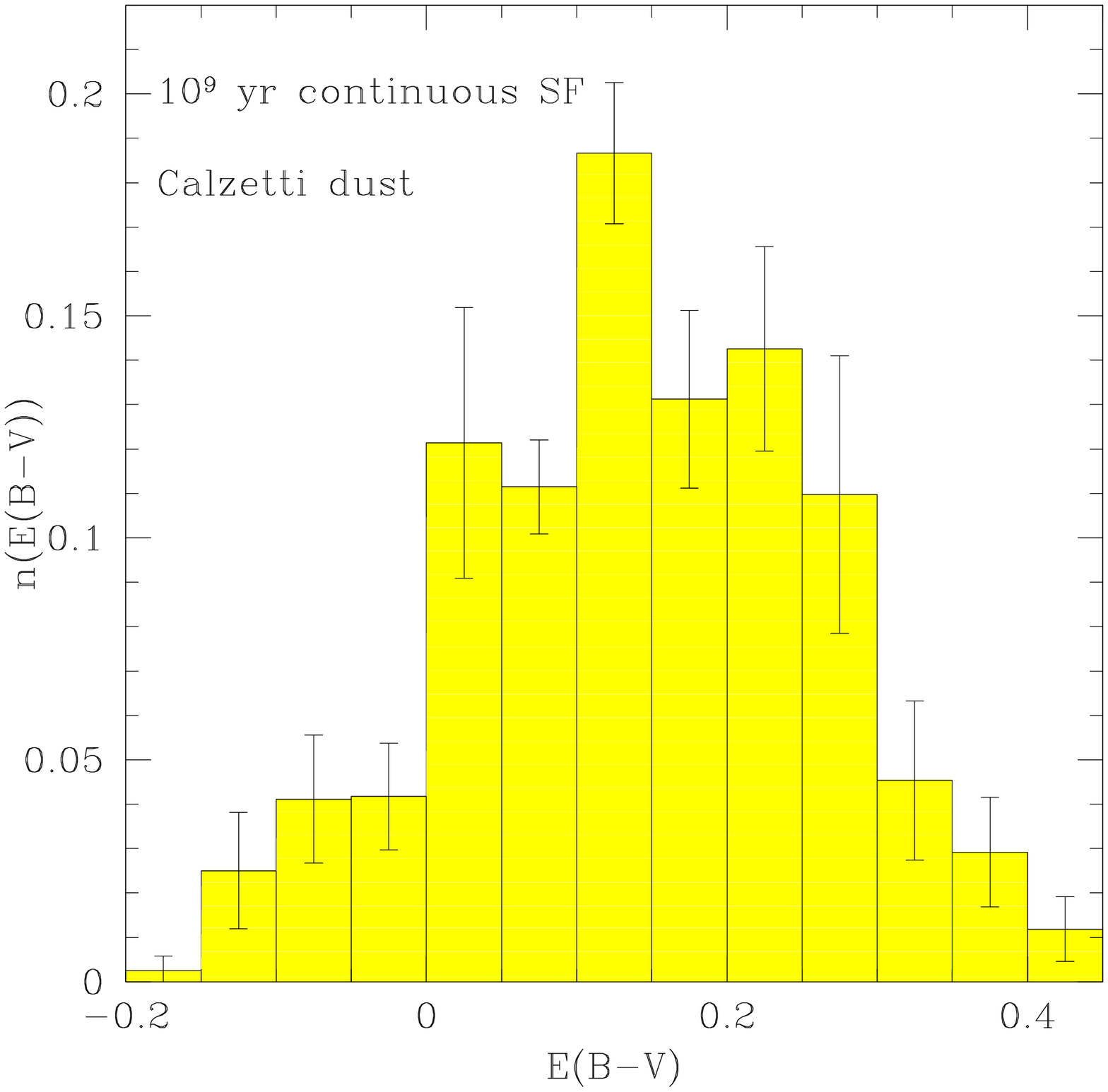,height=7cm}}
\caption[]{(Left) The observed far--UV luminosity function of $z \sim 3$ Lyman break
galaxies. The HDF points have been re--evaluated using
newly determined effective volumes for the selection criteria outlined
in \cite{D98}], and no re--normalization has been necessary to make the two data sets
consistent with one another. The data are plotted as a function of apparent magnitude, rather
than absolute magnitude-- the observed ${\cal R}$ magnitude corresponds to
$\sim 1700$ \AA\ in the rest frame at $z=3$. (Right) The inferred distribution of 
reddening for the $z \sim 3$ LBGs, assuming that all differences in the far--UV
continua are produced by dust reddening with the reddening relation of \cite{Calzetti97}.
}
\end{figure}

Of course, the observed far--UV luminosity function has almost certainly been
significantly altered by extinction. Any estimate of this effect is likely to
be highly uncertain until it is possible to measure the same galaxies in the
far--UV and in the far--IR (see \S 2.5), but the simulations mentioned 
above allow the construction of an internally--consistent model that is also
consistent with external observations, as discussed in the next section.

\subsection{Colors, Reddening, and Extinction}

As shown in Figure 2b, a by--product of our large spectroscopic sample and 
incompleteness corrections is an estimate 
of the intrinsic distribution of continuum colors among the $z \sim 3$ LBGs. Here 
E(B-V) was simply used as a parameterization of the range of continuum colors
encountered; the negative values of E(B-V) correspond to rare objects which have
very strong line emission in the $G$ band, which can have as much as a 0.2
magnitude effect on the colors--none of the objects in the sample has {\it continuum}
colors bluer than our assumed model spectrum against which we are
measuring the implied reddening. The values of E(B-V) are dependent on our
assumptions about the intrinsic LBG spectrum shape, and on our choice of reddening
law, which in this case is taken from \cite{Calzetti97}. As can be seen from
the figure, most of the LBGs have implied reddening that lies in the range
E(B-V)$=0-0.3$, with a median value of 0.15, which corresponds to an implied
extinction at rest--frame 1500 \AA\ of $\sim 1.7$ magnitudes, or a factor
of $\sim 5$, with a net correction in total UV luminosity across the population in the
spectroscopic sample of a factor of $\sim 7$ (cf. \cite{D98}). 
Individual (rare) objects, however, have implied extinction of
up to $\sim 5$ magnitudes. The implications are that a ``typical'' LBG, with an apparent
magnitude corresponding to m$^{\ast}$ in the luminosity function shown in Figure 2a, and the
median extinction correction, would have a star formation rate of about $65{\rm h}_{50}^{-2}$ 
M$_{\sun}$ yr$^{-1}$
for $\Omega_M=1$; the most luminous objects in the sample, after nominal correction for
extinction, would have SFR in excess of 1000 M$_{\sun}$ yr$^{-1}$ for the same cosmology. 

Within our spectroscopic sample, we also see an {\it observed}
color--magnitude relation, in which the more apparently luminous are also redder, on average. Thus
many of the brightest objects in the sample become significantly brighter when the extinction
corrections are applied, while the faintest objects tend to be bluer and to have smaller
implied extinction corrections. This trend is consistent with what is observed for local galaxies
(cf. \cite{Heckman}). The effects on the luminosity function are explored by \cite{A99}.  

How reliable are these extinction estimates? 
The UV spectral slope has been observed to be a reasonably reliable predictor of
reddening for nearby UV--selected objects for which far--IR observations are also available \cite{Meurer97},
and the implied corrections based on the UV slope are quite consistent with those estimated
from the Balmer emission lines for the few $z \sim 3$ galaxies with near--IR spectroscopy
\cite{P98}. In addition, the typical implied reddening of the $z \sim 3$ LBGs is also
consistent with that estimated for lower redshift objects for which both rest--UV and H$\alpha$ observations
have been compared \cite{Glazebrook98,Tresse97}, after accounting for the difference in rest--UV
wavelength at which the extinction is estimated. Hence, on empirical grounds, the levels
of extinction implied seem reasonable. Nevertheless, the extinction in any individual case
is probably quite uncertain.  

These issues are discussed in much more detail in \cite{S99,A99,D98,Calzetti99,P98}. 

\subsection{Clustering}

\begin{figure}
\centering\mbox{\psfig{file=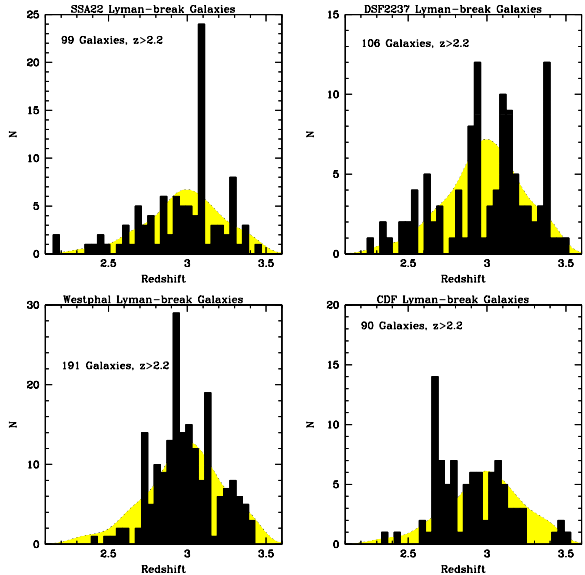,height=8cm}}
\caption[]{Redshift histograms in 4 of our LBG survey fields. The lightly shaded 
histograms represent the overall redshift selection function of the
survey, normalized to the observed number of galaxies in each field.
 }
\end{figure}

Most of what we know at present on the clustering properties of the LBGs has already
been published in a series of papers \cite{S98a,S98b,G98,A98,G99}.
The basic result is that at $z \sim 3$, the relatively bright
LBGs in the ground--based sample are very strongly clustered, with
a co--moving correlation length $r_0$ that is equal to or greater than that of
present--day galaxies. A 
comparison of the clustering strength with the expected clustering of the dark
matter distribution at the same epoch indicates a significant level of bias,
ranging from $b \sim 2$ to $b \sim 6$ depending on the model. Moreover, 
there is evidence for ``clustering segregation'' as a function of UV luminosity,
in the sense that fainter samples of LBGs are less strongly clustered \cite{G99}.
A very good match to the number density and clustering properties of LBGs 
and dark matter halos in simple analytic theory (e.g., \cite{A98, MMW98}) is found
if the shape of the power spectrum is that which best accounts for present--day large
scale structure, normalized by the abundance of rich clusters locally.
Within the context of most of the currently popular models, the characteristic mass of LBGs
in the ground--based samples is $\sim 10^{12}$ M$_{\sun}$; the high bias of the LBGs is
explained by the fact that they reside in halos that are rare at $z \sim 3$. A large number
of theoretical papers have addressed the clustering of LBGs at high redshift 
(e.g.,\cite{Mo96,Jing98,Bagla98,
Governato98,Coles98,Katz98,MMW98,Wechsler98,SPF98,Kauffmann98}),
essentially all finding the observations consistent with general hierarchical
models if the LBGs are tracing the most massive virialized halos at $z \sim 3$.

In \cite{A98} we point out that the agreement of the dark matter models and
the observations of real galaxies (based upon UV luminosity) in terms of abundance and clustering
strength {\it requires} that
there be a monotonic relationship between UV luminosity and dark matter halo mass, with
relatively small scatter. In \cite{A99} we consider the details of the implications
for the relationship between mass and luminosity for LBGs given the observations and
simple assumptions about the dark matter distribution. 
 
It is likely that the largest peaks in redshift histograms such as those shown in Figure
4 are the progenitors of rich clusters of galaxies today \cite{S98a, Governato98}, and are seen
prior to collapse. Thus, many of the LBGs' descendents are likely to be found in rich
environments today. It is plausible that LBGs are closely related to massive galaxies
in the universe today, and that the star formation
one sees at $z \sim 3$ is producing the stars presently found in the bulges of early type
spirals and in ellipticals. As always, a direct connection to present--day galaxies
is not straightforward, and is model--dependent. At present, there are significant
differences among the various groups in the detailed model predictions (e.g. \cite{Baugh98,Kauffmann98,SPF98}); 
the differences are almost entirely due to the star formation recipes, and not
to the behavior of the dark matter.

\subsection{Kinematics}

One of the most important pieces of information that is still missing but
that is observationally tractable is the dynamical masses of LBGs. As seen
above, the clustering properties already suggest association with large
masses, in the context of any reasonable hierarchical structure formation
model. However, as discussed in \cite{P98,A98}, the actual masses of
the halos that are as abundant as observed LBGs vary considerably as
a function of $\Omega_M$ for a fixed power spectrum shape.  A critical
test of the validity of any particular structure formation scenario would
be actual direct measurements of LBG masses. 

The far--UV spectra of LBGs, while very interesting, are severely
limited for measuring any kind of dynamical information because
most of the absorption lines are produced by outflowing interstellar
gas, and Lyman $\alpha$ emission is severely affected by radiative
transfer effects and by external absorption. It is essential, therefore,
to use spectral diagnostics that are more likely to reflect gravitationally
induced motions.


Toward this end, we have attempted pilot observations targeting the
nebular lines in the rest--frame optical (observed near--IR)(\cite{P98})
in an effort to measure line widths as a crude estimate of mass. 
At present the observations are so difficult
that the progress is painstakingly slow and the
results are somewhat ambiguous from a dynamical point of view. The interpretation
of line widths as indicative of masses is also fraught with uncertainty; observed line
widths are essentially always an underestimate of the true circular velocity of
a galaxy because most of the emission emerges from a small region that is
often on the rising part of the rotation curve-- see also \cite{MMW98}. 
Nevertheless, we plan a major
program of near--IR spectroscopy of LBGs in the near future, using 8m--class
telescopes. At the very least, the data can be used as a cross--check on the
extinction corrections discussed above (cf. \cite{P98}), and, by measuring
[OIII], H$\beta$, and [OII], estimates of chemical abundances will be feasible. 

\subsection{LBGs at Other Wavelengths}

An obvious and important extension of the issues discussed in \S 2.2 above is
to attempt to observe LBGs at rest--frame far--IR wavelengths, as both
a cross-check on extinction estimates, and as a means of determining how
much overlap there is between the LBG population discovered in the UV
and the luminous IR galaxies discovered using SCUBA, discussed extensively
at this meeting. The main problem is that few of the LBGs, even after
extinction correction, imply star formation rates that would be detectable
with SCUBA, where the detection/confusion limit is typically $>400$ M$_{\sun}$ yr$^{-1}$ for the adopted
cosmology.  The first results on pointed sub--mm observations of LBGs 
will soon be available \cite{Chapman99}. On the other hand, given the extinction
estimates described in \S 2.2, a considerable fraction of the far--IR background
must be produced by the equivalent of LBGs at intermediate to high redshifts (cf. \cite{Calzetti99}). 

In addition, we have been pursuing a program of near--IR imaging of
a substantial subset of the ground--based LBG sample, and excellent near--IR
data for LBGs in the HDF will soon be available (e.g., \cite{D99}). These observations
will improve the wavelength baseline for reddening estimates, and offer the possibility
of crude assessments of the stellar populations in these early star--forming galaxies.

In general, multi--wavelength campaigns will be crucial for understanding the complete
energetics and nature of LBGs. 

\section{Beyond $z \sim 3$}

Within the past couple of years, inferences on the global star formation 
history over $\sim 90\%$ of the age of the universe have become possible (cf., e.g., \cite{Madau96,MPD98,
Sawicki97}. The first estimates \cite{Madau96} were based upon a combination of spectroscopic
redshift surveys \cite{Lilly96}, photometric redshifts in the HDF \cite{Connolly97}, and
photometric Lyman break galaxies in the HDF. Quite rightly, these first results have
motivated both a great deal of excitement, and a large number of papers pointing out
how the picture could very well be misleading. There are two obvious problems, one of
which has been very much in the forefront at this meeting: the obvious one is that
the HDF results only account
for galaxies that are selected in the far--UV, and do not account for the ``optically dark''
component of galaxy formation (e.g. \cite{Blain98}). 
We have seen above, and elsewhere, that even for galaxies
selected in the UV, most of the energy is apparently absorbed and is presumably re--emitted
in the far--IR. A second possible problem is that the HDF is a very small area, and 
it is quite possibly dangerous to reach universal conclusions based on a 2\arcm\ patch of
sky, no matter how exquisite the data. While it will take considerable time to fully
address the first problem, the second one is in a regime that can be addressed directly
using large--area ground--based surveys. 

One of the most intriguing results from the initial forays into understanding the
``star formation history of the universe'' was the apparently rapid decline in the
star formation density for redshifts beyond $z \sim 3$ \cite{Madau96,MPD98}. This result
was qualitatively consistent with the decline observed for quasars at both optical
\cite{Schmidt95,Kennefick95} and radio \cite{Shaver98} wavelengths, and provided the
rather satisfactory feeling that perhaps one was observing the entire epoch of galaxy
formation, all the way to the beginning of the ``dark ages''. 

\begin{figure}
\centering\mbox{\psfig{file=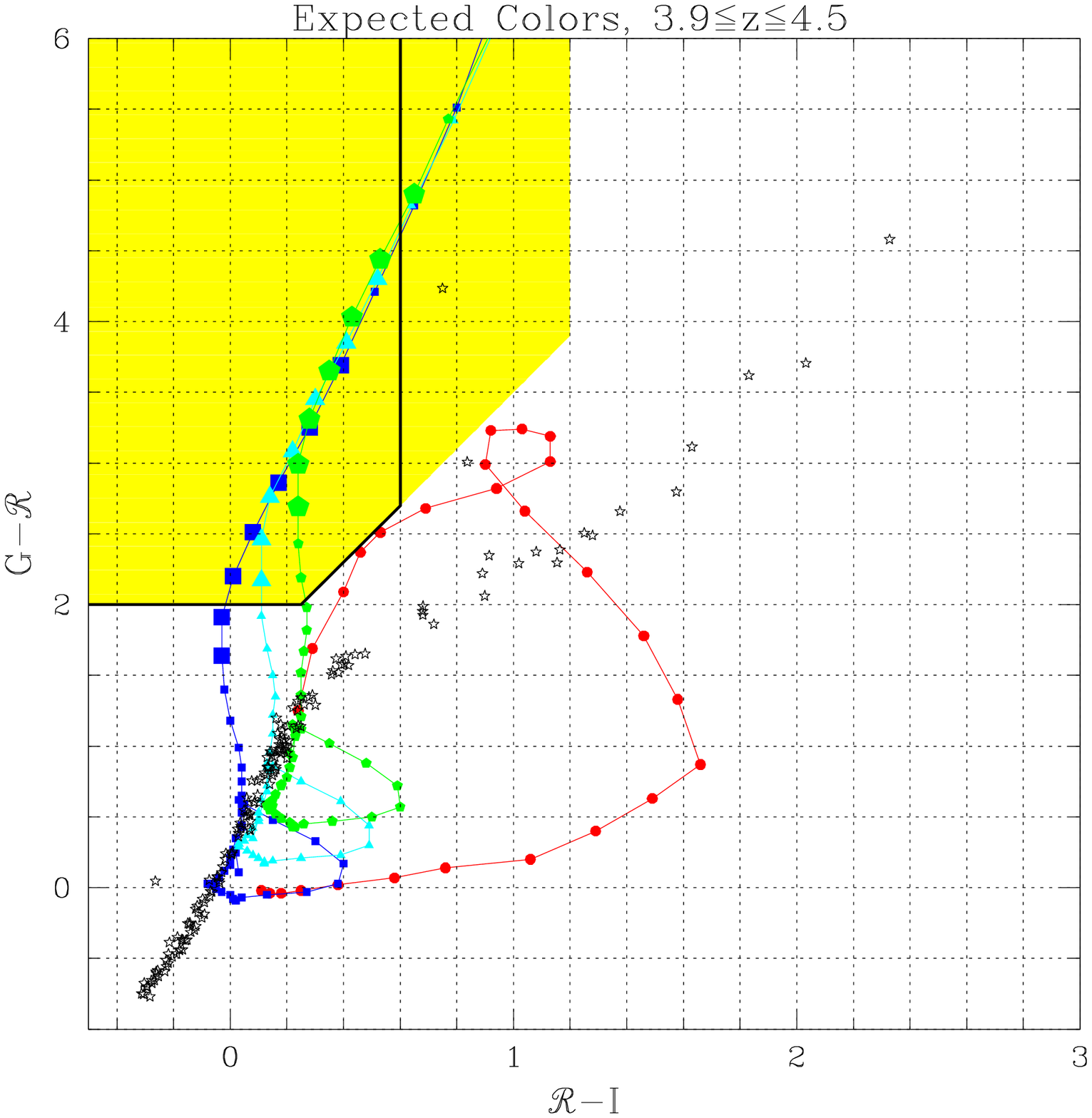,height=7cm}\psfig{file=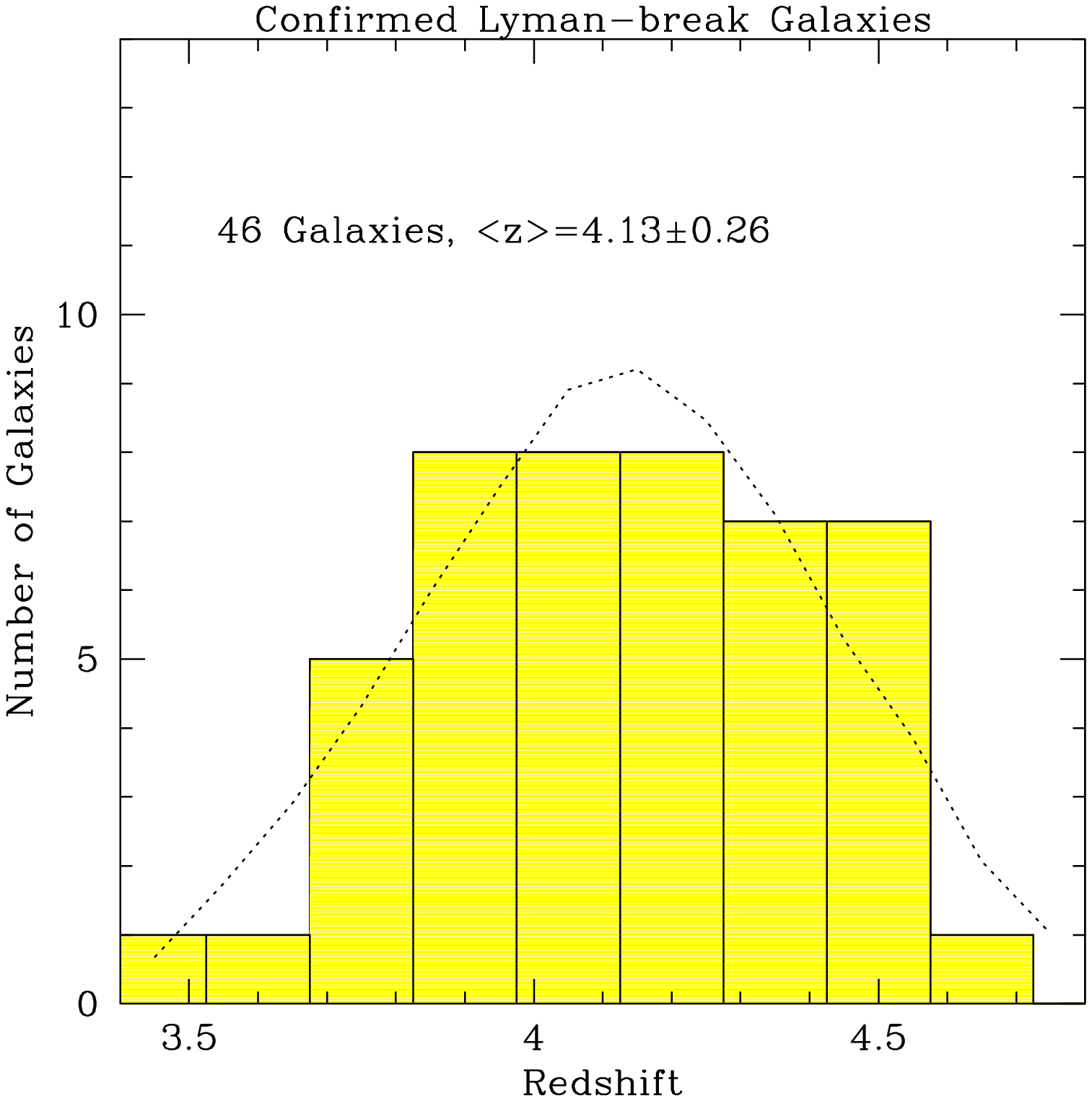,height=7cm}}
\caption[]{(Left) A plot similar to Figure 1a, showing the selection criteria used for
the $z \sim 4$ LBG survey, and the color tracks for model galaxies
having E(B-V)=0,0.15, and 0.30 (the large symbols denote galaxies in the
range $3.9 \le z \le 4.5$). Note that one expects to have some contamination from
intermediate redshift early--type galaxies (tracks running from $z=0.1$ to $z=2.5$ are
shown with solid dots). 
(Right) The current redshift histogram for the $z \sim 4$ sample. The dotted curve represents
the model predictions of the redshift distribution for galaxies having properties that are the
same as those observed at $z \sim 3$, subject to the modeled incompleteness. 
}
\end{figure}

Given the very solid baseline of our $z \sim 3$ photometric and spectroscopic samples, 
we set out to assemble an equivalent and analogous sample at $z \sim 4$ in order 
to test the validity of the HDF result on the evolution of the UV luminosity density
with redshift. Figure 4a illustrates the color criteria used to select the same type
of objects present in the $z \sim 3$ sample, in order to make the cleanest possible
differential comparison. The results of our preliminary spectroscopic follow--up
are shown in Figure 4b. A full description of the survey and results are
given in \cite{S99}.  

The spectroscopic results so far indicate that
the population of UV--luminous galaxies at $z \sim 4$ is very similar to that
observed at $z \sim 3$ -- while we do not have as much basis for measuring the UV
continua of the $z \simgt 4$ galaxies, the assumption that they have the same
distribution of intrinsic colors as measured for the $z \sim 3$ population 
results in the predicted redshift distribution shown in Figure 4b, which is
so far quite consistent with the spectroscopic redshifts. The spectroscopy, while
not as extensive as for the $z \sim 3$ sample, allows a robust estimate of the effective
volume of the survey using techniques analogous to those described above for the 
$z \sim 3$ sample, and also allows a correction for contamination by lower--redshift
interlopers, which is $\sim 20\%$ for the color criteria shown in Figure 4a. 
When we then compare the far--UV luminosity functions at $z \sim 3$ and $z \sim 4$,
the result is shown in Figure 5a. Integrating to obtain 
UV luminosity density for the range of intrinsic luminosity in common for the
two samples, the result is $\rho_{UV}(z=3)/\rho_{UV}(z=4)=1.1\pm0.3$,
independent of cosmology \cite{S99}. {\it Thus, at least at the bright end of the far--UV
luminosity function, there is no significant difference in the luminosity distribution or
in the normalization between $z \sim 3$ and $z \sim 4$}. This result is purely empirical
and depends on few assumptions.

\begin{figure}
\centering\mbox{\psfig{file=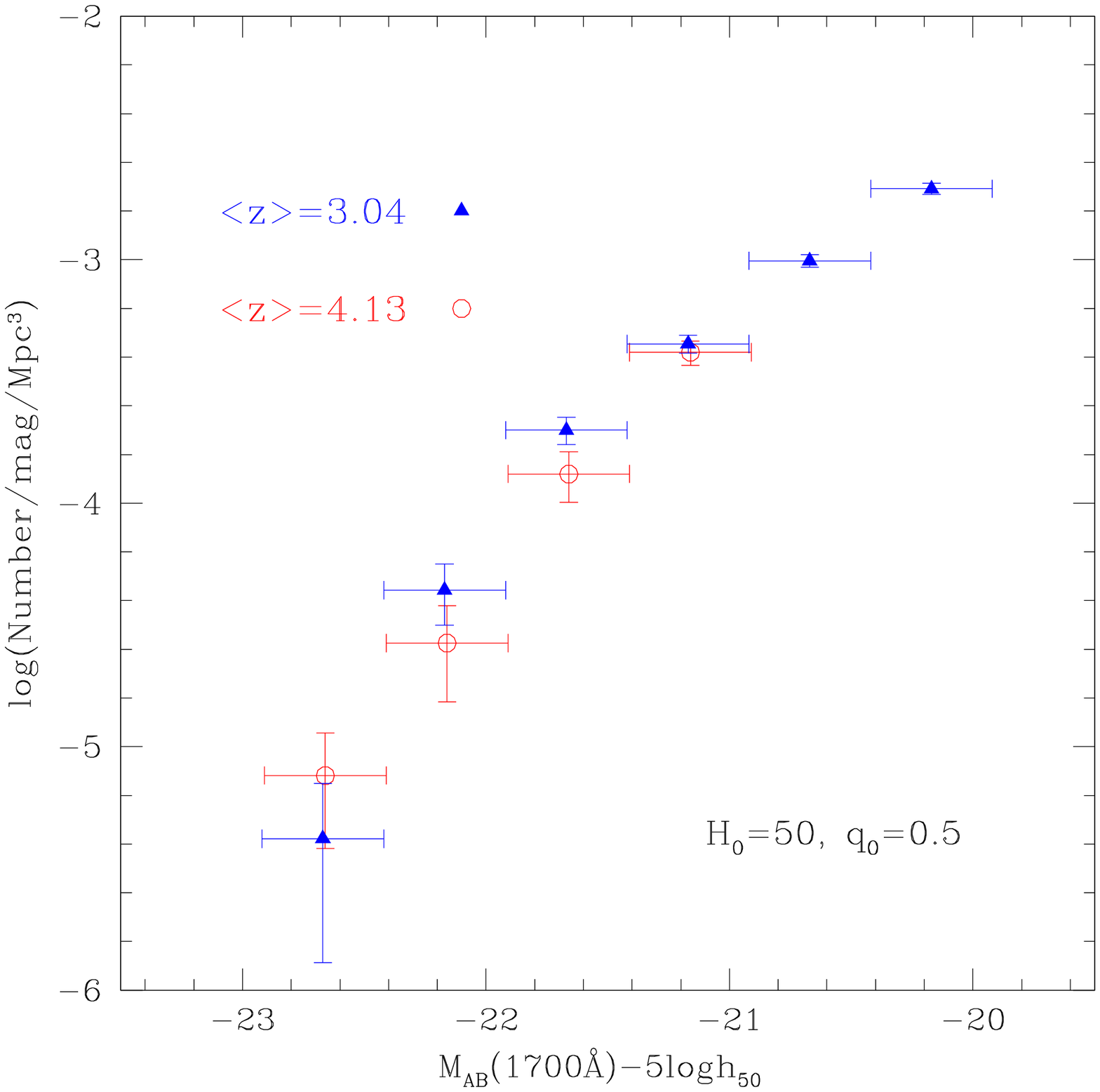,height=8cm}\psfig{file=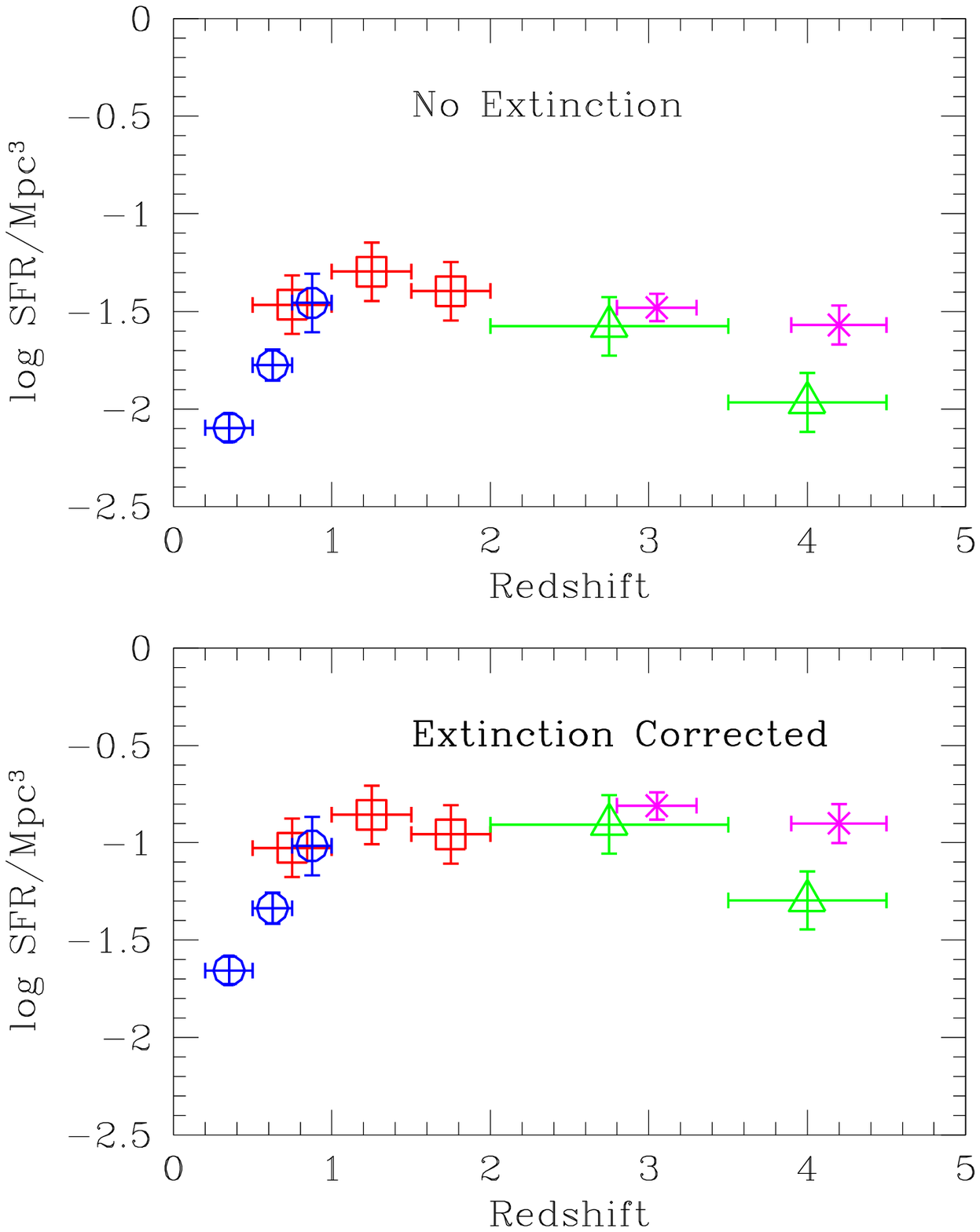,height=8cm}}
\caption[]{(Left) A comparison of the bright ends of the $z \sim 4$ and the $z \sim 3$
luminosity functions for LBGs. Lower $\Omega_m$ cosmologies make the $z \sim 4$
distribution slightly brighter relative to that at $z \sim 3$. Note that both the
shape and normalization of the luminosity distributions are very similar. (Right) The
revised versions of the ``star formation history of the Universe'' from UV luminosity
density measurements. See text for discussion.}
\end{figure}

Of obvious interest (albeit subject to many more uncertainties), 
given the discussions at this meeting, is how these new surveys covering
volumes $\sim 200$ times larger than that of the HDF affect the inferences about the
star formation history of the universe. In Figure 5b, we have plotted the new results
together with earlier results, where in all cases we have integrated over the luminosity
functions only down to the equivalent of $\sim 0.1$L$^{\ast}$ for consistency (this is an extrapolation
for all but the $z\sim3$ LBG sample). Integrating the fitted LFs to arbitrarily small
luminosities would increase the high redshift points by another factor of $\sim 1.7$
relative to the lower redshift points because of the much steeper LF that applies. 
The extinction--corrected version was obtained by  
conservatively applying the median extinction correction discussed above for the
$z \sim 3$ and $z \sim 4$ samples. The point we have tried to make \cite{S99} in this
exercise is that the HDF may have provided misleading results, probably due to sample
variance given the small volume, and at this point the much better constraints provided
by the large ground--based samples are quite consistent with a universe in which the
co--moving SFR density remained constant between $z \sim 4.5$ and $z \sim 1$. 
This could have many implications, which space constraints prevent us from discussing here;
some are discussed in \cite{S99}. 

\section{Summary}

LBGs offer the distinct advantage of allowing wholesale surveys of galaxies
in the distant universe, where it is straightforward to choose nearly volume
limited samples of star forming galaxies in a prescribed redshift range, and where
follow--up work is within reach of the current generation of 8m--class telescopes
and current instruments. 
The efficiency of the Lyman break technique has allowed the first robust estimates of the
clustering properties of distant galaxies, and one of the most robust
luminosity functions determined for any sample beyond the local universe.
The clustering properties in particular have allowed for direct testing
of general frameworks for galaxy formation, and the good statistics allow
for detailed comparison with models of galaxy and structure formation, creating a new
interface with theory that is certain to yield progress in the coming few years.

On the other hand, the LBG surveys are possibly a very incomplete picture of what is going
on at high redshift. The technique almost certainly misses where most of the
energy produced by star formation at high redshift is emerging (although one can
attempt to account for at least some of this with extinction estimates, and
it is not yet clear how many {\it objects} are being missed), and it
is not yet known if the overall picture painted by LBGs for the history
of star formation parallels that which will eventually emerge from studies
of high redshift galaxies in the sub-mm regime. It is also not possible at present
to make more than plausibility arguments connecting LBGs to present--day galaxies,
and the technique is largely insensitive to star formation which has occurred 
prior to the epoch of observation. 

A sensible and constructive combination of
what we are all learning using all of the various techniques available should nicely
fill in a scenario of which we now have only a thumbnail sketch. 
What is really amazing
is how far things have come in just the last few years for understanding the nature
of high redshift galaxies. With all of the capabilities becoming available in the
near and not too distant future, the prospects for imminent progress are immense.   
 
\acknowledgements{We are grateful to all of the people at all of the observatories
who have made possible the observations described above. CCS thanks the organizers
for the invitation to an very stimulating meeting, and for their patience in
awaiting this written contribution.} 


%
\vfill

\begin{bloisbib}
\bibitem{A98} Adelberger, K.L., Steidel, C.C., Giavalisco, M., Dickinson, M.,
Pettini, M., \& Kellogg, M. 1998, ApJ, 505, 18
\bibitem{A99} Adelberger, K.L., et al 1999, in preparation
\bibitem{Bagla98} Bagla, J.S. 1998, MNRAS, 297, 251
\bibitem{Baugh98} Baugh, C.M., Cole, S., Frenk, C.S., \& Lacey, C.G. 1998, ApJ, 498, 504
\bibitem{Blain98} Blain, A., Smail, I., Ivison, R., \& Kneib, J-P. 1998, MNRAS, in press 
(astro-ph/9806062)
\bibitem{Calzetti97} Calzetti, D. 1997, AJ, 113, 162 
\bibitem{Calzetti99} Calzetti, D., and Heckman, T. 1999, ApJ, in press (astro-ph/9811099)
\bibitem{Chapman99} Chapman, S. et al. 1999, in preparation.
\bibitem{Coles98} Coles, P., Lucchin, F., Mattarese, S., \& Moscardini, L. 1998, MNRAS, in press
(astro-ph/9803197)
\bibitem{Cohen96} Cohen, J.G., Hogg, D.W., Pahre, M.A., and Blandford, R.D. 1996, ApJ, 462, L9
\bibitem{Connolly97} Connolly, A.J., Szalay, A.S., Dickinson, M, SubbaRao, M.U., \& Brunner, R.J. 1997,
ApJ, 486, L11
\bibitem{Cowie96} Cowie, L.L., Songaila, A., Hu, E.M., and Cohen, J.G. 1996, AJ, 112, 839
\bibitem{D98} Dickinson, M. 1998, in The Hubble Deep Field, eds. M. Livio, S.M. Fall,
\& P. Madau 1998, (Cambridge: Cambridge University Press), in press (astro-ph/9802064)
\bibitem{D99} Dickinson, M. et al. 1999, in preparation
\bibitem{Ellis96} Ellis, R.S., Colless, M., Broadhurst, T., Heyl, J., and Glazebrook, K. 1996, MNRAS, 279, 47
\bibitem{G98} Giavalisco, M., Steidel, C. C., Adelberger, K. L., Dickinson, M. E., Pettini, M., \& Kellogg, M. 1998a, ApJ, 503, 543 
\bibitem{G99} Giavalisco, M. et al. 1999, ApJ, in preparation.
\bibitem{Glazebrook98} Glazebrook, K. Blake, C., \& Economou, F. 1998, MNRAS, submitted
\bibitem{Governato98} Governato, F., Baugh, C. M., Frenk, C. S., Cole, S., Lacey, C. G., Quinn, T., \& Stadel, J. 1998,
Nature, 392, 359 
\bibitem{Heckman} Heckman, T.M. 1998, in {\it The Most Distant Radio Galaxies}, in press (astro-ph/9801155)
\bibitem{Jing98} Jing, Y.P., \& Suto, Y. 1998, ApJ, 494, L5
\bibitem{Katz98} Katz, N.S., Hernquist, L., \& Weinberg, D.H. 1998, ApJ, submitted (astro-ph/9806257)
\bibitem{Kauffmann98} Kauffmann, G., Colberg, J.M., Diaferio,  A., and White, S.D.M. 1998, MNRAS,
submitted (astro-ph/9809168)
\bibitem{Kennefick95} Kennefick, J.D., Djorgovski, S.G., \& de Carvalho, R. R. 1995, AJ, 110, 2553
\bibitem{Lilly96} Lilly, S.J., LeF\`evre, O., Hammer, F., \& Crampton, D. 1996, ApJ, 460, L1
\bibitem{Lowenthal97} Lowenthal, J.D., et al. 1997, ApJ, 481, 673
\bibitem{MPD98} Madau, P., Pozzetti, L., \& Dickinson, M.E. 1998, ApJ, 498, 106 (MPD)
\bibitem{Madau96} Madau, P., Ferguson, H.C., Dickinson, M., Giavalisco, M., Steidel, C.C.,
\& Fruchter, A. 1996, MNRAS, 283, 1388
114, 54
\bibitem{Meier76} Meier, D.M. 1976, ApJ, 203, L103
\bibitem{Meurer97} Meurer, G.R., Heckman, T.M., Lehnert, M.D., Leitherer, C., and Lowenthal, J. 1997, AJ,
\bibitem{MMW98} Mo, H., Mao, S., \& White, S.D.M. 1998, MNRAS, submitted (astro-ph/9807341) 
\bibitem{Mo96} Mo, H..J., and Fukugita, M. 1996, ApJ, 467, L9
\bibitem{Oke95} Oke, J. B. et al. 1995, PASP 107, 3750
\bibitem{P98} Pettini, M., Kellogg, M., Steidel, C. C., Dickinson, M., Adelberger, K. L., \& Giavalisco, M. 1998a, ApJ, in
press (astro-ph/9806219)
\bibitem{P97} Pettini, M., Steidel, C.C., Adelberger, K.L., Kellogg, M., Dickinson, M., and Giavalisco, M.
1997, in {\it ORIGINS}, eds. J.M. Shull, C.E. Woodward, and H. Thronson, (San Francisco: ASP)
(astro-ph/9708117) 
\bibitem{Sawicki97} Sawicki, M. J., Lin, H., \& Yee, H. K. C. 1997, AJ, 113, 1
\bibitem{Schech76} Schechter, P. 1976, ApJ, 203, 297
\bibitem{Schmidt95} Schmidt, M., Schneider, D.P., \& Gunn, J.E. 1995, AJ, 110, 68
\bibitem{Shaver98} Shaver, P., Hook, I.M., Jackson, C.A., Wall, J.V., \& Kellermann, K.I. 1998,
in Highly Redshifted Radio Lines, eds. C. Carilli, S. Radford, K Menten, \& G. Langston 
(astro-ph/9801211)
\bibitem{SPF98} Somerville, R.S., Primack, J.R., \& Faber, S.M. 1998, MNRAS, submitted
(astro-ph/9806228)
\bibitem{S99} Steidel, C.C., Adelberger, K.L., Giavalisco, M., Dickinson, M., and Pettini, M. 1999,
ApJ, submitted (astro-ph/9811399)
\bibitem{S98a} Steidel, C.C., Adelberger, K. L., Dickinson, M., Giavalisco, M.,
Pettini, M. \& Kellogg, M. 1998a, ApJ, 492, 428
\bibitem{S98b} Steidel, C.C., Adelberger, K. L., Giavalisco, M., Dickinson, M.,
Pettini, M. \& Kellogg, M. 1998b, Phil.Trans. R.S., in press (astro-ph/9805267).
\bibitem{S96} Steidel, C.C., Giavalisco, M., Pettini, M., Dickinson, M., \&
Adelberger, K. L. 1996a, ApJ, 462, L17
\bibitem{SPH} Steidel, C.C., Pettini, M., and Hamilton, D. 1995, AJ, 110, 2519
\bibitem{Tresse97} Tresse, L., \& Maddox, S.J. 1998, ApJ, 495, 691
\bibitem{Wechsler98} Wechsler, R. H., Gross, M. A. K., Primack, J. R.,
Blumenthal, G. R. \& Dekel, A. 1998, ApJ, submitted (astro-ph/9712141)
\end{bloisbib}
\end{document}